\documentclass[a4paper,twocolumn,9pt,3p,times]{elsarticle}
\usepackage{amsmath,amsthm,amssymb,amsthm,dsfont,mathrsfs,mathtools}
\usepackage[colorlinks]{hyperref}
\usepackage[usenames,dvipsnames,svgnames,table]{xcolor}
\usepackage[english]{babel}
\usepackage{tikz,tkz-berge,tkz-graph,pgfplotstable}
\usetikzlibrary{positioning,chains,fit,shapes,calc,backgrounds,external}

\journal{Physics Letters A}

\begin{document}
\begin{frontmatter}
\title{On the connection between linear combination of entropies and linear combination of extremizing distributions}
\author[cbpf]{Gabriele Sicuro}\ead{sicuro@cbpf.br}
\author[cbpf]{Debarshee Bagchi}\ead{debarshee@cbpf.br}
\author[nist,sf]{Constantino Tsallis}\ead{tsallis@cbpf.br}
\address[cbpf]{Centro Brasileiro de Pesquisas F\'isicas, Rua Dr. Xavier Sigaud 150, 22290-180, Rio de Janeiro, Brazil}\address[nist]{Centro Brasileiro de Pesquisas F\'isicas and National Institute of Science and Technology for Complex Systems,\\  Rua Dr. Xavier Sigaud 150, 22290-180, Rio de Janeiro, Brazil}
\address[sf]{Santa Fe Institute, 1399 Hyde Park Road, Santa Fe, NM, 87501 USA}

\begin{abstract}
We analyze the distribution that extremizes a linear combination of the Boltzmann--Gibbs entropy and the nonadditive $q$-entropy. We show that this distribution can be expressed in terms of a Lambert function. Both the entropic functional and the extremizing distribution can be associated with a nonlinear Fokker--Planck equation obtained from a master equation with nonlinear transition rates. Also, we evaluate the entropy extremized by a linear combination of a Gaussian distribution (which extremizes the Boltzmann--Gibbs entropy) and a $q$-Gaussian distribution (which extremizes the $q$-entropy). We give its explicit expression for $q=0$, and discuss the other cases numerically. The entropy that we obtain can be expressed, for $q=0$, in terms of Lambert functions, and exhibits a discontinuity in the second derivative for all values of $q<1$. The entire discussion is closely related to recent results for type-II superconductors and for the statistics of the standard map.
\end{abstract}


\end{frontmatter}

\section{Introduction}
Generalized Fokker--Planck (FP) equations have been successfully employed to describe anomalous diffusion processes in a plethora of different contexts \cite{chavanis2008}. Diverse phenomena, like diffusion in porous media \cite{muskat1938,berkowitz1997,klemm1997}, surface growth in fractals \cite{spohn1993}, black hole radiation \cite{kaniadakis2005}, diffusion in optical lattices \cite{lutz2003,douglas2006}, heartbeat of healthy individuals \cite{peng1993}, disordered superconductors \cite{zapperi2001,andrade10}, financial indices \cite{borland2002,boon2005}, among others, cannot be described in terms of linear FP equations \cite{risken1984}. All these complex systems are characterized by the presence of long-range space and/or time correlations. In order to deal with such systems, two main modifications of the linear FP equations have been proposed and are being worked upon, namely, fractional FP equations with nonlocal operators \cite{metzler2000}, and nonlinear FP equations with nonlinear transition rates \cite{frank2005}. In the latter case,  $q$-statistics \cite{tsallis1988,tsallis2009,tsallis2009b,tsallis2009} has provided useful tools for the study of the entropy associated with nonlinear FP \cite{plastino1995, tsallis1996, fuentes2008,combe2015}. 
In particular, in a recent study on the thermostatistics of the overdamped motion of interacting particles, a nonlinear FP equation associated with a linear combination of entropies has been used to describe the diffusion of vortices in type-II superconductors \cite{andrade10}. On the other hand, in a numerical work on the nonlinear dynamics of the standard map \cite{tirnakli15}, a linear combination of a Gaussian and a $q$-Gaussian distribution has been successfully adopted as a fitting ansatz. A linear combination of a Gaussian and a $q$--Gaussian appears also in the model proposed by \citet{miah2014} for the velocity distribution of tracer particles in quantum turbulence. In their model, tracer particles spend some time in a normal liquid and in a superfluid alternately. The $q$--Gaussian distribution emerges from the superfluid dynamic, whereas the Gaussian is related to the normal regime: the resulting distribution is therefore a linear combination of the two distributions. Remarkably, their model reproduces the experimental data very well \cite{miah2014}.

Motivated by these recent works, in the present paper, we systematically explore the connection between linear combination of two entropies and linear combination of the probability distributions that optimize each of these extropies separately. Although a linear combination of two entropies is not maximized by a linear combination of the two corresponding extremizing distributions, still in some cases this assumption can be adopted, within a certain error, as a working ansatz.

In order to do this, we employ the following approach. In Section \ref{S1} we construct a nonlinear FP equation from a master equation, assuming nonlinear transition rates. In Section \ref{S2} we show that the introduced FP equation can be associated to an entropy which is a linear combination of the Boltzmann--Gibbs (BG) entropy and the nonadditive $q$-entropy. The exact stationary state solution of such a FP equation was found recently in terms of the Lambert $W$ function, which in two limits yields the Gaussian and the $q$-Gaussian distributions \cite{andrade10,casas2015}. Finally, in Section \ref{S3}  we consider a linear combination of these two probability distributions, namely a Gaussian and a $q$-Gaussian, that individually extremize the (additive) BG entropy and the nonadditive $q$-entropy respectively. We reconstruct the entropy corresponding to the linear combination of the probability distributions using the prescription presented by \citet{plastino14}. We obtain the exact expression of the entropy for the case $q=0$, and we discuss numerically the other cases.

\section{A nonlinear Fokker--Planck equation}\label{S1}
A FP equation for a ``walker'' (i.e., the state of our system) in one dimension can be derived from a master equation on a one dimensional regular lattice with step size $\epsilon\to 0^+$ \cite{gardiner85} 
\begin{equation}
\frac{\partial P(n,t)}{\partial t}=\sum_{m=-\infty}^{+\infty} P(m,t)w(n|m;t)-P(n,t)\sum_{m=-\infty}^{+\infty}w(m|n;t).
\end{equation}
In the previous expression, $P(n,t)$ is the probability of finding the walker in the position $x=n\epsilon$, $n$ integer number, at the time $t$ and $w(m|n;t)$ is the transition rate from the site $x=n\epsilon$ to the site $x=m\epsilon$ at the time $t$. In the $\epsilon\to 0$ limit, introducing a properly rescaled probability density $p(x,t)\propto\frac{1}{\epsilon}P(n,t)$, it is well known that a FP--type equation is obtained, the coefficients of which depend on the form assumed by the transition rate $w(m|n;t)$. In particular, if
\begin{equation}
w(m|n;t)\equiv w(m|n)=-\frac{f(n\epsilon)}{\epsilon}\delta_{n,m+1}+D\frac{\delta_{n,m+1}+\delta_{n,m-1}}{\epsilon^2},
\end{equation}
with \begin{equation}f(x)=-\frac{d\phi(x)}{d x}\end{equation}
being the \textit{external force} on the walker, $\phi$ the external confining potential, and $D>0$ the \textit{diffusion constant}, the linear FP equation
\begin{equation}
\frac{\partial p(x,t)}{\partial t}=-\frac{\partial}{\partial x}\left[f(x)p(x,t)\right]+D\frac{\partial^2 p(x,t)}{\partial x^2}\label{fpe}
\end{equation}
is recovered in the $\epsilon\to 0$ limit taking $x\equiv n\epsilon$ fixed. 

Eq.~\eqref{fpe} is however inadequate for the description of nonlinear diffusion processes in which the transition rate can depend on a certain power of $P$. Indeed, the diffusion process of a specific walker takes place, in general, in presence of a large number of other diffusing walkers and it may happen, therefore, that $w(m|n;t)$ depends also on the probability that the arrival site $m$ is already occupied. A very simple assumption is
\begin{multline}
w(m|n;t)=-\frac{f(n\epsilon)}{\epsilon}\delta_{n,m+1}\\+\left[a+b\frac{ P^{\nu-1}(n,t)+P^{\nu-1}(m,t)}{2\epsilon^{\nu-1}}\right]\frac{\delta_{n,m+1}+\delta_{n,m-1}}{\epsilon^2},
\end{multline}
in which we have introduced a nonlinear diffusion term. Here $a,b$ are real, positive constants and we assume $\nu\geq 1$. We can put $b\equiv 1$ without loss of generality. In the $\epsilon\to 0$ limit, we obtain the following \textit{generalized FP equation} for the probability density distribution $p$,
\begin{equation}
\frac{\partial p(x,t)}{\partial t}=-\frac{\partial}{\partial x}\left[f(x)p(x,t)\right]
+a\frac{\partial^2p(x,t)}{\partial x^2}+\frac{1}{\nu}\frac{\partial^2 p^{\nu}(x,t)}{\partial x^2}.
\label{fpenl}
\end{equation}
The equation for the stationary distribution $p(x)$ is
\begin{multline}
\frac{a+ p^{\nu-1}(x)}{p(x)}\frac{\partial p(x)}{\partial x}=f(x),\\\text{and assuming }\lim_{x\to\pm\infty}p(x)=\lim_{x\to\pm\infty}\frac{\partial p(x)}{\partial x}=0,
\end{multline}
the solution of which can be written as
\begin{equation}
p(x)=\left[a W\left(\frac{p_0^{\nu-1}}{a}e^{\frac{p_0^{\nu-1}-(\nu-1)(\phi(x)-\phi_0)}{a}}\right)\right]^{\frac{1}{\nu-1}}.\label{stsol}
\end{equation}
Here $\phi_0\equiv\phi(0)$, $p_0\equiv p(0)$ is to be fixed by imposing normalization, and $W(z)$ is the Lambert function, defined as the solution of the equation $W(z)e^{W(z)}=z$. \citet{andrade10} showed that the nonlinear FP equation \eqref{fpenl} with $\nu=2$ describes properly vortices in type-II superconductors; they also derived the stationary distribution in Eq.~\eqref{stsol} for this particular case. Eq.~\eqref{fpenl} has been recently obtained also by \citet{casas2015} in the analysis of the nonlinear Ehrenfest model. Observe that
\begin{equation}
p(x)\xrightarrow{a\gg 1}p_0\, e^{-\frac{\phi(x)-\phi_0}{a}},
\end{equation}
thus recovering the BG distribution. The previous formula shows that $a\propto T$, i.e., $a$ plays the role of a temperature. 
In contrast, by taking the $a\to 0$ limit (low temperature limit), we have
\begin{equation}
p(x)\xrightarrow{a\to 0}p_0 \,\exp_{2-\nu}\left({-\frac{\phi(x)-\phi_0}{p_0^{\nu-1}}}\right),\label{qG}
\end{equation}
where we have introduced the \textit{$q$--exponential function}
\begin{equation}
\exp_q(x)\coloneqq\begin{cases}\exp(x)&\text{if $q=1$,}\\\left[1+(1-q)x\right]_+^\frac{1}{1-q}&\text{if $q\neq 1$,}\end{cases}
\end{equation}
and $[x]_+\coloneqq x\,\theta(x)$. The equation obtained for $a=0$ and its  solutions were analyzed by \citet{plastino1995} in the study of diffusion processes in porous media, and later associated to a Langevin--type equation by \citet{borland1998}. Its derivation from a master equation is discussed in \cite{curado03,nobre2004}. Finally, let us mention that Eq.~\eqref{fpenl} can also be discussed for $\nu<1$, see \cite{pedron2005,rodriguez2014}.

\section{Entropy and stationary solution: from linear combination of entropies to the extremizing distribution}\label{S2}
\citet{schwammle07} showed that we can associate an entropic functional $S[p]$ to Eq.~\eqref{fpenl} under the assumption that the entropy is trace-form
\begin{equation}\label{entropic}
S[p]\coloneqq\int_{-\infty}^{+\infty}s(p(x,t))\,dx,\quad s(0)=s(1)=0,\quad \frac{d^2s(p)}{dp^2}\leq 0.
\end{equation}
Here and in the following we will suppose that the space of possible states is the real line. For the sake of completeness, we sketch their main results here. Let us consider a generic nonlinear FP equation of the form
\begin{equation}\label{GenFPE}
\frac{\partial p(x,t)}{\partial t}=-\frac{\partial}{\partial x}\left[f(x)p(x,t)\right]+\frac{\partial}{\partial x}\left[\Omega[p(x,t)]\frac{\partial p(x,t)}{\partial x}\right].
\end{equation}
In the previous expression, $f(x)\equiv -\frac{d\phi(x)}{dx}$ as before, whilst $\Omega[p]\geq 0$ is a certain function of the distribution $p$. To preserve the norm, we require that 
\[\lim_{x\to\pm\infty}p(x,t)=\lim_{x\to\pm\infty}f(x)p(x,t)=\lim_{x\to\pm\infty}\frac{\partial p(x,t)}{\partial x}=0, \ \forall t.\]
Let us now associate to the previous equation the generic trace-form entropy in Eq.~\eqref{entropic} and the free-energy functional
\begin{equation}
F[p]\coloneqq\int_{-\infty}^\infty\phi(x) p(x,t)d x-\frac{S[p]}{\beta},\quad\beta>0 \,.
\end{equation}
Imposing that the $H$--theorem holds, i.e. $\frac{\partial F}{\partial t}\leq 0$, we obtain
\begin{equation}\label{Hth}
\Omega[p]=-\frac{p}{\beta}\frac{d^2 s(p)}{dp^2}.
\end{equation}
If we now consider Eq.~\eqref{fpenl}, Eq.~\eqref{Hth} becomes
\begin{equation}
-\frac{1}{\beta}\frac{d^2 s(p)}{dp^2}=\frac{a+p^{\nu-1}}{p}.
\end{equation}
The constraints $s(0)=s(1)=0$ fix the form of $s(p)$ for $\nu>1$ as
\begin{equation}\label{sentropy1}
s(p)=ak\beta\left(- p\ln p+\frac{p-p^{\nu}}{a\nu(\nu-1)}\right),
\end{equation}
where $k$ is a positive constant. Observe that $s(p)$ is concave and has a maximum in the interval $[0,1]$ for \[p_\text{max}=\left[a W\left(\frac{1}{a} e^{\frac{1}{a(\nu-1)}-\nu+1}\right)\right]^\frac{1}{\nu-1}.\] The entropy associated with the nonlinear FP equation \eqref{fpenl} can therefore be expressed as a \textit{linear combination} of a BG entropy and a \textit{nonadditive $q$-entropy} with entropic index $\nu$:
\begin{equation}
 S[p]=a\beta S_{\text{BG}}[p]+\frac{\beta}{\nu}S_{\nu}[p].\label{entropy} 
\end{equation}
The nonadditive entropy $S_q[p]$ is defined as follows \cite{tsallis1988}:
\begin{equation}
S_q[p]\coloneqq\frac{1-\int_{-\infty}^{+\infty} p^q(x)\,d x}{q-1}.
\end{equation}
We can evaluate the corresponding distribution by extremizing the entropy functional \eqref{entropy} with the constraints $\int p(x)\,d x=1$ and $\int (\phi(x)-\phi_0)p(x)\,d x=u$ internal energy. Let us write the following Lagrangian functional:
\begin{multline}
\mathcal L[p]=S_{\text{BG}}[p]+\frac{S_{\nu}[p]}{a\nu}\\-\gamma_1\left[\int p(x)\,d x-1\right]\\-\gamma_2\left[\int (\phi(x)-\phi_0)p(x)\,d x-u\right],
\end{multline}
where suitable Lagrange multipliers have been introduced. Computing $\frac{\delta\mathcal L}{\delta p}=0$ it is easily seen that the distribution maximizing the entropic functional \eqref{entropy} has the form of the stationary solution \eqref{stsol}:
\begin{equation}
p(x)=\left[a W\left(\frac{p_0^{\nu-1}}{a} e^{\frac{p_0^{\nu-1}}{a}-\gamma_2(\nu-1)(\phi(x)-\phi_0)}\right)\right]^{\frac{1}{\nu-1}}\,,
\label{stsol2}
\end{equation}
where $p_0$ is fixed by imposing the normalization condition.

\begin{figure}\centering
\pgfplotsset{scaled y ticks=false,scaled x ticks=false,every axis/.append style={legend style={mark size=2pt}}}
\includegraphics{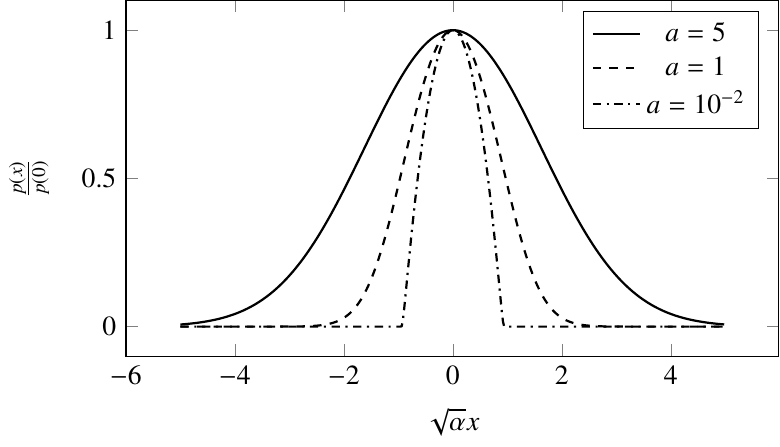}
\caption{Stationary solution of Eq.~\eqref{stsol3} for different values of $a$ and $\nu=2-q=2$.}\label{F1}
\end{figure}

By comparison with Eq.~\eqref{stsol}, if we impose that the stationary solution is the extremizing distribution for the functional \eqref{entropy}, we have that
\begin{equation}
\gamma_2\equiv\frac{1}{a}.
\end{equation}
Let us from now on consider a harmonic potential
\begin{equation}
\phi(x)=\alpha x^2,\quad \alpha>0.
\end{equation}
Eq.~\eqref{stsol2} can be written as
\begin{equation}
p^{(2-\nu)}_a(x)=\left[a\,W\left(\frac{p_0^{\nu-1}}{a}\exp\left(\frac{p_0^{\nu-1}-\alpha (\nu-1) x^2}{a}\right)\right)\right]^{\frac{1}{\nu-1}}.\label{stsol3}
\end{equation}
Interestingly enough, for $a\gg 1$, the distribution is a Gaussian one, the linear diffusive term being dominant. Instead, for $a\to 0$, we recover a $q$-Gaussian shape with $q=2-\nu<1$. On the other hand, if $q=\nu=1$ the behavior is purely Gaussian for all values of $a$. The distribution, for $\nu=2$, is plotted in Fig.~\ref{F1} for different values of $a$: 
as anticipated above, this particular case appeared in the study of the diffusion of $N$ vortices in a type-II two-dimensional superconductor of size $L_x\times L_y$, $L_y\ll L_x$ in \cite{andrade10}. Observe that the distribution \eqref{stsol3} maximizes the linear combination of entropies in Eq.~\eqref{entropy} and, obviously, it is \textit{not} merely a linear combination of the optimizing distributions of the two entropies.

\section{The inverse road: from linear combination of distributions to the corresponding entropy}\label{S3}
Let us search now for a trace-form entropic functional $\tilde S[p]=\int_{-\infty}^{+\infty} \tilde s[p(x)]dx$ having a linear combination of a Gaussian and a $q$-Gaussian as extremizing distribution, i.e., a distribution given by a linear combination of the probability distributions extremizing the BG entropy and the $S_q$ entropy respectively.  As anticipated, a fitting ansatz in this form was recently used by \citet{tirnakli15} in the numerical investigation of the statistics of the standard map, and it appears also as solution of a model for turbulence \cite{miah2014}. In both models, however, it is not clear if any entropic functional $\tilde S[p]$ could be associated to this limiting probability distribution function. We address this question here.

To derive the required entropic functional, let us proceed in generality. Let us suppose that a certain entropic functional in the form
\begin{equation}
\Sigma[p]=\int_{-\infty}^{+\infty} \sigma(p(x))dx
\end{equation}
is given. From the maximum entropy principle, imposing the constraints $\int\epsilon(x)p(x)dx=u$ and $\int p(x)dx=1$, we obtain the equation
\begin{equation}
\frac{d\sigma(p)}{dp}-\gamma_1-\gamma_2\epsilon=0,
\end{equation}
where $\epsilon$ plays the role of an energy function, and $\gamma_1$ and $\gamma_2$ are two Lagrange multipliers to be determined. Let us assume that the previous equation is solved by the distribution $P=P(\epsilon)$, with $P'(\epsilon)<0$ in its domain of definition\footnote{Here we assume that the domain of $P$ as function of $\epsilon$ does not coincide necessarily with the set of physically acceptable values of $\epsilon$, but with the values of $\epsilon$ such that $P(\epsilon)$ is a well defined quantity.}. We assume also that there exists a value $\bar\epsilon\in\mathds R$ such that $P\colon [\bar\epsilon,+\infty)\to(0,1]$. In particular, we have $P(\bar\epsilon)=1$. Under these assumptions, the function $\epsilon_P(p)\coloneqq P^{-1}(p)$ is well defined in $(0,1]$. Then\footnote{Notice that any monotonic function of the obtained trace-form entropy is extremized, for the same set of constraints, by the same distribution.}
\begin{equation}
\sigma(p)=\gamma\left(\frac{\int_0^p\epsilon_P(y)\,dy}{\int_0^1\epsilon_P(y)\,dy}-p\right),\label{sentropy2}
\end{equation}
for some constant $\gamma$, where we imposed $\lim_{p\to 0}\sigma(p)=\lim_{p\to 1}\sigma(p)=0$. Moreover, the condition $\gamma \geq 0$ must be satisfied to guarantee concavity. Applying Eq.~\eqref{sentropy2} to the distribution in Eq.~\eqref{stsol2} with $\epsilon(x)\equiv\phi(x)-\phi_0$, we recover the linear combination of a BG entropy and a $S_q$ entropy. 

Let us now evaluate the entropic functional $\Sigma$ on a discrete probability distribution $P=\{p_i\}_{i=1,\dots,W}$, $W\in\mathds N_0$. We have
\begin{equation}
    \Sigma[P]\coloneqq\sum_{i=1}^W\sigma(p_i).\label{entropydis}
\end{equation}
The entropic functional in Eq.~\eqref{entropydis} satisfies the first three Khinchin's axioms \cite{khinchin1957}. Indeed, under our hypotheses,
\begin{itemize}
    \item $\Sigma[P]$ is a continuous function of its arguments;
    \item $\Sigma[P]$ is maximized by the uniform distribution;
    \item adding a zero-probability state, the entropy does not change, being $\sigma(0)=0$ (expansibility).
\end{itemize}
The fourth Khinchin's axiom is violated, unless $\Sigma$ is exactly the Boltzmann--Gibbs entropy. The \textit{composability property} \cite{tempesta2016} holds only for a specific two-parameter form of $\sigma(p)$ \cite{tempesta2015,sicuro2016} and, therefore, the entropy $\Sigma$ is not composable in general. Our result in Eq.~\eqref{sentropy2} is consistent with the uniqueness result obtained by \citet{naudts2008} on generalized exponential families. A more general recipe for the construction of a trace form entropy optimized by a given energy density distribution function is discussed by \citet{naudts2008} and \citet{plastino14}.

Observe that similar distributions optimize similar entropic functionals. Indeed, let us assume that two invertible probability distribution $P_1(\epsilon)$ and $P_2(\epsilon)=P_1(\epsilon)+\delta\rho(\epsilon)$ are given, $\delta\ll 1$. Let us introduce now
\begin{align}
    \epsilon_1(p)\coloneqq& P_1^{-1}(p),\\ \epsilon_2(p)\coloneqq& P_2^{-1}(p)=\epsilon_1(p)-\delta\frac{\rho\left(\epsilon_1(p)\right)}{P_1'\left(\epsilon_1(p)\right)}+o(\delta).
\end{align}
If we require
\begin{equation}
    \sup_{0\leq p\leq 1}\left|\int_{\epsilon_1(0)}^{\epsilon_1(p)}\rho(y)\,dy \right|= K<+\infty,\quad K\in\mathds R^+,
\end{equation}
the result follows from Eq.~\eqref{sentropy2} observing that
\begin{equation}
\left|\frac{\int_0^p\epsilon_1(y)\,dy}{\int_0^1\epsilon_1(y)\,dy}-\frac{\int_0^p\epsilon_2(y)\,dy}{\int_0^1\epsilon_2(y)\,dy}\right|\leq \tilde K \delta
\end{equation}
for some constant $\tilde K$.

\begin{figure*}\centering
\pgfplotsset{scaled y ticks=false,scaled x ticks=false,every axis/.append style={legend style={mark size=2pt}}}
\includegraphics{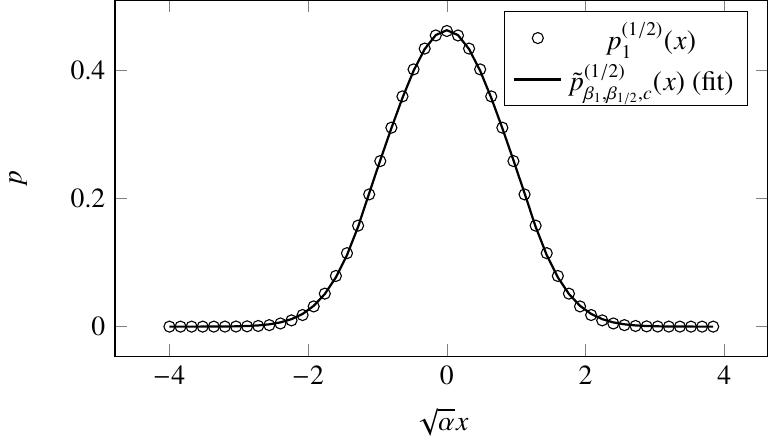}\quad
\includegraphics{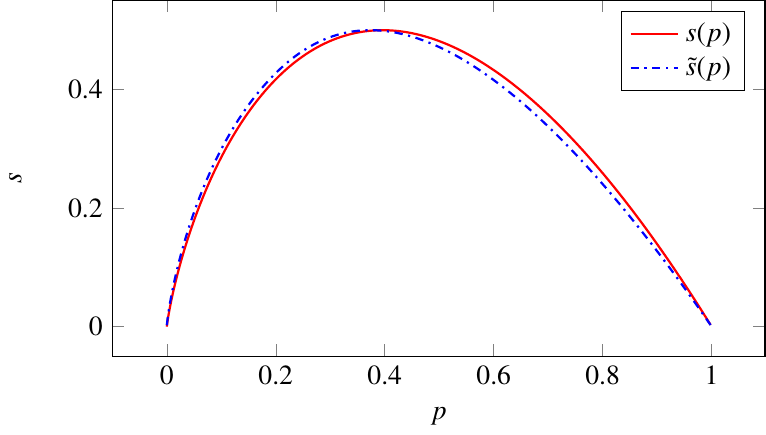}
\caption{On the left, plot of $p^{(1/2)}_1(x)$ and of the function $\tilde p^{(1/2)}_{\beta_1,\beta_q,c}(x)$ obtained as best fit of $p^{(1/2)}_1(x)$, with $\beta_1,\beta_q,c$ as free parameters. Both functions are rescaled in such a way that their maximum is $\frac{1}{2}$. On the right, plot of $s(p)$ and $\tilde s(p)$.}\label{F2}
\end{figure*}

Let us consider now the following probability distribution function
\begin{multline}
\tilde p_{\beta_1,\beta_q,c}^{(q)}(\epsilon)=\frac{1-c}{Z_1}\exp\left(-\beta_1\epsilon\right)+\frac{c}{Z_q}\exp_q\left(-\beta_q \epsilon\right),\\ c\in[0,1],\quad q\in(-\infty,1].\label{lcdis}
\end{multline}
 Here $Z_1$ and $Z_q$ are normalizing constants, depending on the expression of $\epsilon=\epsilon(x)$. Observe that the function above has a discontinuity in the first derivative for $\epsilon=\frac{1}{\beta_q}\frac{1}{1-q}$ and therefore we expect that it cannot be a solution of a regular FP equation. We can evaluate the quantity \eqref{sentropy2} by inverting the equation above. Obviously, the distribution in Eq.~\eqref{lcdis} is \textit{not} the optimizing distribution of the linear combination of entropies in $S[p]$ appearing in Eq.~\eqref{entropy} for any value of its parameters, except in the trivial case $q=1$. Remarkably enough, the distribution in Eq.~\eqref{lcdis}, if used as a fitting ansatz, can approximate quite well the exact solution \eqref{stsol}, this approximation becoming trivially exact for $q=1$. Let us call now $\tilde S[p]=\int_{-\infty}^{+\infty} \tilde s(p(x))dx$ the entropic functional associated to $\tilde p_{\beta_1,\beta_q,c}^{(q)}$, obtained using the procedure sketched above. If we tune the parameters of $\tilde p_{\beta_1,\beta_q,c}^{(q)}$ to get the best fit of $p^{(q)}_a(x)$, for a fixed value of $q$, $a$ and $\alpha$, the two entropic forms $S[p]$ and $\tilde S[p]$ are different but still very similar, as expected. As particular case, in Fig.~\ref{F2} we consider $\epsilon(x)=x^2$ and the distribution in Eq.~\eqref{stsol3} obtained for $a=1$ and $\nu=\frac{3}{2}$, \[p^{(1/2)}_1(x)=\left[W\left(\sqrt{p_0}\exp\left(\sqrt{p_0}-\frac{x^2}{2}\right)\right)\right]^2.\]
We plot also the function $\tilde p^{(1/2)}_{\beta_1,\beta_q,c}(x)$ obtained as best fit of $p^{(1/2)}_1(x)$, with $\beta_1,\beta_q,c$ as free parameters, and the functions $s(p)$ and $\tilde s(p)$. The plot of $\tilde s(p)$ is obtained numerically using the parameters of the best fit. As expected, the entropic forms reconstructed using our method do not differ very much. Therefore a linear combination of an $S_q$ entropy and a Boltzmann--Gibbs entropy can be a good approximation to the true entropy maximized by a linear combination of a $q$-Gaussian and a Gaussian, and \textit{vice versa}.

The $q=0$ case can be treated analytically. In the notation above, we have
\begin{equation}
\epsilon_P(p)=\begin{cases}\begin{split}
&-\frac{1}{\beta_1}\ln\frac{Z_1p}{1-c}\\&\qquad 0\leq p\leq \min\{p_c,1\}\end{split}\\
\begin{split}&\frac{1}{\beta_0}-\frac{Z_0}{c\beta_0}p+\frac{1}{\beta_1}W\left[\frac{Z_0\beta_1}{Z_1\beta_0}\frac{1-c}{c}e^{\frac{Z_0\beta_1}{c\beta_0}\left(p-\frac{c}{Z_0}\right)}\right]\\&\qquad\min\{p_c,1\}\leq p\leq 1.\end{split}\end{cases}
\end{equation}
Here we have introduced
\begin{equation}
p_c\coloneqq \frac{1-c}{Z_1}e^{-\frac{\beta_1}{\beta_0}}.
\end{equation}
Let us suppose now that $p_c\in(0,1)$. Firstly, let us compute
\begin{equation}
    \mathcal C(p)\coloneqq \int_0^p\epsilon_{p^(0)_{\beta_1,\beta_q,c}}(y)\,d y.
\end{equation}
For $p\in[0,p_c]$ we have
\begin{equation}
    \mathcal C(p)= \int_0^p\left[-\frac{1}{\beta_1}\ln\frac{Z_1y}{1-c}\right]\,d y=\frac{p}{\beta_1}-\frac{p}{\beta_1}\ln \frac{Z_1 p}{1-c}.
\end{equation}
We have then that, for $p_c\in(0,1)$ and $p\in(p_c,1]$,,
\begin{multline}\allowdisplaybreaks
    \mathcal C(p)=\mathcal C(p_c)\\\allowdisplaybreaks+\int^p_{p_c}\left\{\frac{1}{\beta_0}-\frac{Z_0}{c\beta_0}y+\frac{1}{\beta_1}W\left[\frac{Z_0\beta_1}{Z_1\beta_0}\frac{1-c}{c}e^{\frac{Z_0\beta_1}{c\beta_0}\left(y-\frac{c}{Z_0}\right)}\right]\right\}\,d y\\\allowdisplaybreaks
    =\mathcal C(p_c)+\frac{p-p_c}{\beta_0}-Z_0\frac{p^2-p_c^2}{2c\beta_0}y\\+\frac{c\beta_0}{2Z_0\beta_1^2}\left\{2W\left[\frac{Z_0\beta_1}{Z_1\beta_0}\frac{1-c}{c}e^{\frac{Z_0\beta_1}{c\beta_0}\left(p-\frac{c}{Z_0}\right)}\right]\right.\\\allowdisplaybreaks
    -2W\left[\frac{Z_0\beta_1}{Z_1\beta_0}\frac{1-c}{c}e^{\frac{Z_0\beta_1}{c\beta_0}\left(p_c-\frac{c}{Z_0}\right)}\right]\\\allowdisplaybreaks
    +W^2\left[\frac{Z_0\beta_1}{Z_1\beta_0}\frac{1-c}{c}e^{\frac{Z_0\beta_1}{c\beta_0}\left(p-\frac{c}{Z_0}\right)}\right]\\\allowdisplaybreaks
    \left.-W^2\left[\frac{Z_0\beta_1}{Z_1\beta_0}\frac{1-c}{c}e^{\frac{Z_0\beta_1}{c\beta_0}\left(p_c-\frac{c}{Z_0}\right)}\right]\right\}.
\end{multline}
In the previous formula, we have used the fact that, for $a,b>0$,
\begin{equation}
    \int_{z_1}^{z_2} W\left(ae^{bx}\right)\,dx=\left.\frac{2W\left(ae^{bx}\right)+W^2\left(ae^{bx}\right)}{2b}\right|_{z_1}^{z_2}.
\end{equation}
It follows immediately that
\begin{equation}\label{spa}
\frac{\tilde s(p)}{\gamma}=-p+\frac{1}{\mathcal C(1)}\begin{cases}\frac{p}{\beta_1}-\frac{p}{\beta_1}\ln \frac{Z_1 p}{1-c}&0<p\leq p_c,\\
\mathcal C(p)&p_c\leq p<1.\end{cases}
\end{equation}
The expression above is quite involved and nontrivial, and it has as extremizing distribution the function given in Eq.~\eqref{lcdis}. Extending the application of our entropic form to the discrete case, in the spirit of Eq.~\eqref{entropydis}, we observe that $\tilde S[P]$ cannot be put in the general form presented in \cite{hanel2011}, despite the fact that it is trace-form and the first three Khinchin axioms are fulfilled. Computing $\tilde S[P]$ on the uniform distribution $P=\left\{\frac{1}{W}\right\}_{i=1,\dots,W}$, we have
\begin{equation}
    \tilde S\left[\left\{\frac{1}{W}\right\}_{i=1,\dots,W}\right]\sim \frac{\ln W}{\beta_1}\qquad \text{for $W\gg 1$}.
\end{equation}

\begin{figure*}\centering
\includegraphics{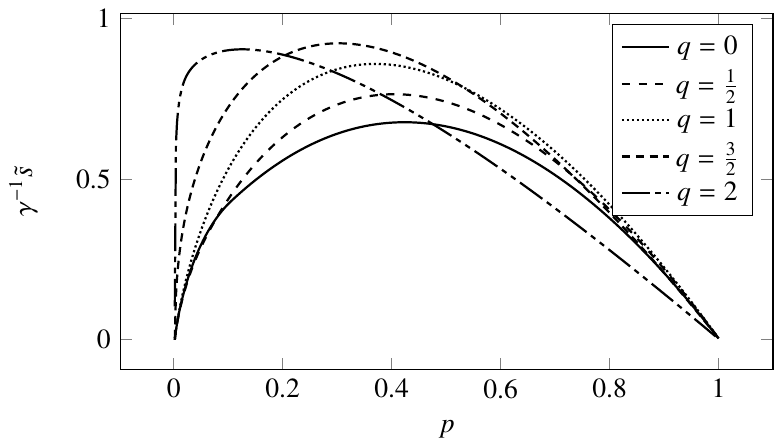}\quad
\includegraphics{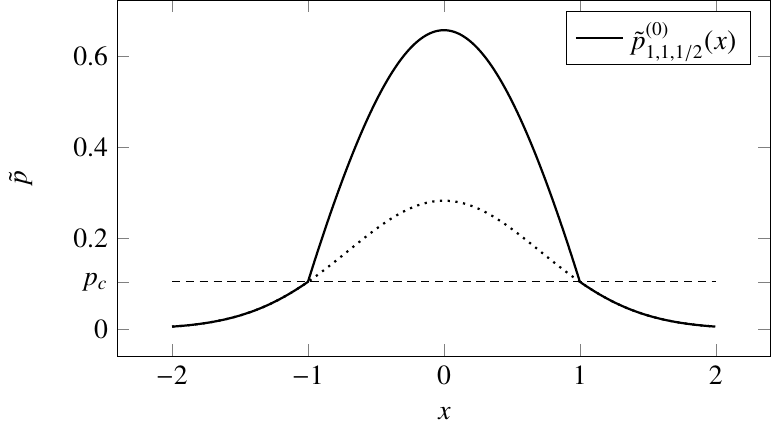}
\caption{On the left, plot of $\tilde s$ for different values of $q$ and $\beta_1=\beta_q=2c=1$ obtained assuming $\epsilon(x)=x^2$. On the right, plot of the optimizing distribution $\tilde p^{(0)}_{1,1,\frac{1}{2}}(x)$: the dotted curve is the Gaussian distribution appearing in the linear combination.}\label{F3}
\end{figure*}
Finally, observe that the quantity $p\frac{d^2\tilde s(p)}{dp^2}$ is not continuous, due to the discontinuity of the first derivative of $\tilde p^{(q)}_{\beta_1,\beta_q,c}(x)$ when $p_c\in(0,1)$. This fact reinforces that the previous quantity cannot be used in a regular FP equation. Moreover, the discontinuity of the second derivative of $\tilde s(p)$ prevents the existence of an escort distribution associated to this entropic form, and the definition of a Fisher information matrix, in the sense specified in \cite{naudts2008}.

In Fig.~\ref{F3} we compare the result in Eq.~\eqref{spa} with other entropic functionals corresponding to different values of $q$. The curves are obtained by integration, as in Eq.~\eqref{sentropy2}, of the (numerically derived) inverse of the probability distribution function 
\begin{multline}
\tilde p_{1,1,\frac{1}{2}}^{(q)}(x)=\frac{e^{-x^2}}{2Z_1}+\frac{e_q^{-x^2}}{2Z_q},\\
Z_q=\begin{cases}\frac{2 \sqrt{\pi} \Gamma\left(\frac{1}{1-q}\right)}{(3-q) \sqrt{1-q} \Gamma\left(\frac{3-q}{2-2q}\right)}& q\in(-\infty,1),\\
\sqrt{\pi}&q=1,\\
\frac{\sqrt{\pi} \Gamma\left(\frac{3-q}{2q-2)}\right)}{\sqrt{q-1} \Gamma\left(\frac{1}{q-1}\right)}&q\in(1,3).\end{cases}\end{multline}

\section{Conclusions}
In the present paper we discussed a FP equation that can be associated to an entropic functional given by a linear combination of the Boltzmann-Gibbs entropy and the nonadditive $q$-entropy. The stationary solution of this equation can be expressed in terms of a Lambert $W$ function, as already pointed out by \citet{andrade10} and \citet{casas2015}. This distribution smoothly interpolates between two limiting distributions, namely, a Gaussian and a $q$-Gaussian. Clearly this solution is \textit{not} a linear combination of a Gaussian and a $q$-Gaussian. Inspired by some numerical results obtained in \cite{tirnakli15} on the standard map, we investigated also the inverse road, trying to reconstruct the entropy associated to the linear combination of a Gaussian and a $q$-Gaussian. In particular, using a procedure introduced recently  in \cite{plastino14}, we were able to reconstruct the analytical expression of the trace-form entropic functional having a linear combination of a Gaussian and a $q$-Gaussian with $q=0$ as extremizing distribution, showing that it can be expressed again in terms of Lambert functions. This entropic functional has a quite involved expression and, moreover, it presents a (jump) discontinuity in the second derivative and. Therefore, it is not possible to construct a proper FP equation associated to it. 

As a closing remark, let us emphasize that, in spite of the conceptual differences that we have exhibited here (namely that entropic extremization and linear combination do {\it not} generically commute), the numerical discrepancies can be, for appropriate choice of the fitting parameters, almost negligible for a wide range of physical situations.

\section*{Acknowledgments}
The authors thank Evaldo M.~F. Curado for useful discussions. The authors acknowledge also the financial support of the John Templeton Foundation.
\section*{References}

\bibliographystyle{elsarticle-num-names}
\bibliography{Biblio.bib}
\end{document}